# Quantum states of macrosystems and entropy

**Maria Polski**

East-West University, Chicago, IL, USA

**Vladimir Skrebnev**

Kazan Federal University, Kazan, Russia

## Abstract

The paper examines and critiques the expression of entropy as the logarithm of the number of quantum states of a physical system. Boltzmann's method of expressing entropy as the logarithm of the number of states of a gas with a given total energy is analyzed. We demonstrate that entropy is the product of subquantum processes and show that entropy is expressed as the ratio of the logarithm of the maximum number of realizations, over the observation period, of a macroscopic system's states with a given total energy, to the number of occurrences of its quantum states over this time.

Email: vskrebnev@gmail.com

## Introduction

Entropy is a fundamental concept tied to processes in physics, chemistry, and informatics. In philosophy, entropy has come to be used as a metaphor to describe the processes related to chaos, order, and evolution. Some historians and philosophers, such as Hegel, have tried to apply the concept of entropy to the analysis of historical processes. Entropy is used in cosmology to characterize the state of the Universe: it is believed that the Universe, like any other closed system, tends to a state with maximum entropy; heat death of the Universe is a hypothetical scenario in which the Universe eventually reaches a state of thermodynamic equilibrium, i.e., a state with maximum entropy.

The concept of entropy migrated from physics to mathematics, information theory, and other fields of science. Therefore, understanding the physical nature of entropy is especially important. It is to the physical nature of entropy that this work is devoted.

## Entropy and the quantum energy levels

In contemporary physics, entropy $S$ is associated with the number $W$ of quantum mechanical states:

$$S = ln\, W \tag{1}$$

It is argued (see, for example, [1]) that all the quantum mechanical energy levels of the system have to be concentrated in an extremely narrow interval in the vicinity of point $E$, where $E$ is the total energy of the system. In this interval, the probability of a system state corresponding to each level is considered the same and is determined by the canonical distribution formula

$$\rho(E_n) = \frac{e^{-\beta E_n}}{\sum_m e^{-\beta E_m}} \tag{2}$$

where $E_n$ equals $E$.

We know that in a macrosystem in equilibrium, its full energy, in other words, simply energy, is

$$E = \sum_n \rho(E_n)\, E_n \tag{3}$$

and

$$\sum_n \rho(E_n) = 1 \tag{4}$$

If one follows [1] and accepts that quantum mechanical levels of energy are concentrated in an extremely narrow interval, then indeed one could consider that the probability of system states in that interval is the same and equals

$$\rho(E) = \frac{e^{-\beta E}}{\sum_m e^{-\beta E_m}}. \tag{5}$$

Then,

$$\sum_n \rho(E_n) = \rho(E)W = 1 \quad (6)$$

where $W$ is the number of levels in the extremely narrow interval under consideration, that is, the full number of quantum mechanical system states. This is the number used in formula (1).

Formula (6) can be rewritten as

$$\rho(E)W = 1 \quad (7)$$

From (7) it follows that

$$ln\, W = -ln\, \rho(E) \quad (8)$$

However, let us explore the vicinity of point *E*.

Depending on temperature, the energy *E* of a macrosystem can have different values within the system's spectrum of energy levels. The spectrum of energy levels itself, however, is independent of *E*. Therefore, the assumption that the system's energy levels concentrate in an extremely narrow interval around point *E* and must be wandering around depending on the value of *E*, has nothing to do with the physical reality. Accordingly, entropy cannot be equal to the logarithm of the number of states that do not exist. To verify the validity of our criticism, it is necessary to carefully read the relevant sections of textbooks, all the time correlating them with the basic tenets of quantum mechanics.

## The true physical nature of entropy

Let us now show that $-ln\, \rho(E)$ does indeed equal the entropy of the system, and what the real physical processes are that underlie this formula.

Let's substitute $E$ instead $E_n$ into formula (2), and we obtain the expression

$$\rho(E) = \frac{e^{-\beta E}}{\sum_m e^{-\beta E_m}}. \quad (9)$$

Then

$$ln\rho(E) = -\beta E \ - lnZ$$

where

$$Z = \ \sum_m e^{-\beta E_m} \quad (10)$$

Taking into account that $\beta = \frac{1}{kT}$., let us rewrite (10) in the form

$$ln\rho(E) = -\frac{E+kTlnZ}{kT} \quad (11)$$

The quantity - $\kappa TlnZ$ is called the Helmholtz free energy and is denoted by the letter $F$. Accordingly, we have:

$$ln\rho(E) = -\frac{E+kTlnZ}{kT} = \ -\frac{E\ -F}{kT} \quad (12)$$

Using the well-known thermodynamic relationship

$$E– F = TS \quad (13)$$

we find

$$-ln\rho(E) = \ \frac{S}{k} \quad (14)$$

Entropy, as a logarithm of some expression, is a dimensionless quantity. In thermodynamic relations, for convenience, this quantity is multiplied by Boltzmann's constant. Formula (14) gives the value of entropy as a dimensionless quantity.

Thus, entropy is expressed through the logarithm of $\rho(E)$, while the generally accepted interpretation of entropy as a logarithm of a number of some quantum states is absolutely unacceptable.

To properly understand the relationship between entropy and the quantum mechanical states of a system, it is necessary to start from the very beginning.

In compliance with quantum mechanics, the function of the system state can be represented as:

$$\psi = \sum_n c_n(t)\psi_n, \quad (15)$$

where $\psi_n$ is the eigenfunction of the system's Hamiltonian,

$$c_n(t) = c_n(0)\exp\left(-\frac{i}{\hbar}E_n t\right). \quad (16)$$

The quantum mechanical average of the system's energy, i.e., its total energy, is

$$E = \sum_n |c_n(t)|^2 E_n. \quad (17)$$

According to quantum mechanics, $|c_n(t)|^2$ determines the probability of energy $E_n$. It is only possible to talk about the probability of a particular state with energy $E_n$ if the system transitions from one state to another. Therefore, $|c_n(t)|^2$ is the probability of the emergence of a system state with energy $E_n$. The emergence of precisely such states is confirmed by the fact that the system absorbs or emits energy at frequencies

$$\omega_{kl} = \frac{E_k - E_l}{\hbar} \quad (18)$$

Quantum mechanics says nothing about these transitions as such and, accordingly, about the processes that cause these transitions. However, it is obvious that these processes, not discussed in quantum mechanics and occurring at the subquantum level of matter organization, cannot help but exist. It is precisely these processes that are responsible for the emergence of system states with energy $E_n$.

Formula (17) can only be used when a very large number of such transitions occur during the observation time. This means that subquantum processes occur extremely quickly, and the time scale to which we are accustomed would be inconvenient for describing them.

Our prerogative to use the categories of probability theory in quantum mechanics and statistical physics means that in these fields subquantum processes manifest themselves as random processes. However, behind every randomness lies a regularity. Decoding this regularity is a task for the future.

Statistical physics also says nothing about the processes that cause system states with energy $E_n$. In our papers [2,3], it was shown for the first time that the canonical distribution and entropy of the system are the outcomes of subquantum processes. In [2,3], the emergence of the system state with energy $E_n$ , aka $E_n$-state, is called the visit of the system to the state with energy $E_n$. The full number of visits over the observation time is denoted as *N*, and the number of visits to a particular $E_n$-state is denoted as $\nu_n$.

Math experts will confirm that the same value of energy *E* is possible with different sets of values $|c_n(t)|^2$ in (17). Subquantum processes occur in accordance with the law of conservation of energy; that is, they conserve the total energy of the system. Each set of values $|c_n(t)|^2$ corresponds to a certain set of visits, which we call "the visit configuration" or simply "configuration". Each configuration can be realized in *P* ways, corresponding to the number of visit permutations:

$$P = \frac{N!}{\nu_1!\nu_2!\ldots\nu_l!\ldots} \qquad (19)$$

In [2,3], it was shown that the largest number *P* corresponds to the configuration in which

$$\nu_n = \frac{Ne^{-\beta E_n}}{\sum_m e^{-\beta E_m}}. \qquad (20)$$

from where we find

$$\rho(E_n) = \frac{\nu_n}{N} = \frac{e^{-\beta E_n}}{\sum_m e^{-\beta E_m}} \qquad (21)$$

that is, the canonical distribution of the probability of system energy $E_n$. Recall that the canonical distribution characterizes the state of thermodynamic equilibrium of a macrosystem.

In [3] we obtained

$$\frac{lnP_{max}}{N} = -\sum_n \rho(E_n)\, ln\rho(E_n) = -ln\rho(E) = S \qquad (22)$$

## Discussion

Formula (22) means that entropy is the ratio of the logarithm of $P_{max}$ to the number of acts $N$ of occurrence during the observation time of all states with energies $E_n$ . Here $P_{max}$ is the number of ways in which the configuration corresponding to the canonical distribution can be realized during that time. We can assume that each way corresponds to a certain state of the macroscopic system. Let us remember that there is a reason for the emergence of these states, and this reason is subquantum processes.

Boltzmann's remarkable work "Uber die Beziehung zwischen dem zweiten Hauptsatze der mechanischen Warmtheorie und der Wahrscheinlichkeitsrechnung den Satzen uber das Warmegleichgewicht" [4] ("On the Relationship between the Second Fundamental Theorem of the Mechanical Theory of Heat and Probability Calculations Regarding the Conditions for Thermal Equilibrium" ) considers the distribution of gas energy among molecules while maintaining the gas's energy *E*. Boltzmann denotes the total number of gas molecules as *N*, and the number of molecules with energy $E_l$ as $\omega_l$.

Each distribution can be realized in different ways, which Boltzmann calls Komplexions. The number of Komplexions in a distribution is equal to the number of permutations of the molecules:

$$P = \frac{N!}{\omega_1!\omega_2!\ldots\omega_l!\ldots} \tag{23}$$

Boltzmann showed that the maximum number of permutations corresponds to the number of particles at the level $E_k$ , equal to (in modern notation):

$$n_k = N\frac{e^{-\beta E_k}}{Z} \tag{24}$$

where $Z = \sum_m e^{-\beta E_m}$

Boltzmann uses the term "a measure of permutation" for the number of permutations of particles in a gas (the number of Komplexions). The logarithm of the maximum value of this measure up to some constant is equated by Boltzmann with the entropy of the gas. That is (omitting the constant),

$$ln\, P_{max} = S \qquad (25)$$

We can consider that the Komplexions characterize the microstates of the gas, and denote their number as $W$. If $P_{max}$ is replaced with $W$, then we can write (in the modern notation):

$$S = k\, ln\, W \qquad (26)$$

In his work [4] Boltzmann used the assumption that the energy values of gas molecules are discrete. He made this assumption before the dawn of quantum mechanics. Boltzmann himself called it fictitious, and yet it is a kind of premonition of quantum theory! The concept of probability in quantum mechanics is natural and necessary. Boltzmann used probability theory to describe the behavior of gas molecules long before the creation of quantum theory, during the era of classical mechanics, in which probabilities have no place.

On September 5, 1906, Boltzmann committed suicide in a hotel in the Italian town of Duino, hanging himself with a window cord. Boltzmann's suicide is attributed to depression caused by the lack of understanding in the physics community of the ideas of statistical physics he was developing.

In our articles [2, 3], we sought the most probable distribution of states of a macrosystem with total energy $E$, these states being caused by subquantum processes. Boltzmann's remarkable work, to which we refer in those articles, was of great assistance in this endeavor.

Following Boltzmann, entropy can generally be defined as

$$S = \frac{ln\, P}{N} \qquad (27)$$

where $P$ is expressed by formula (19). In this case, the energy of the system can be expressed by the formula (17). By representing the probability of the energy value as

$$|c_n(t)|^2 = \frac{\nu_n}{N}$$

and having performed calculations similar to those performed in [3], we obtained the formula

$$\frac{\ln P}{N} = S = -\sum_n |c_n|^2 \ln |c_n|^2 \quad (28)$$

The irreversible evolution caused by subquantum processes leads the macrosystem to a state of thermodynamic equilibrium with a maximum value of $P$, that is, to a maximum of entropy.

## Conclusion

We have shown that entropy is expressed by formula (22) and formula (28). How formula (1) with its bewildering interpretation of the value $W$ slipped into quantum physics remains a mystery to us. Perhaps it was driven by a desire to emulate Boltzmann and to get a formula that looked like his.

We hope that this work of ours, as well as articles [2,3], will help the reader better understand the physical meaning of entropy, one of the most important characteristics of the phenomena in the world around us.